\documentclass[a4paper,14pt]{extarticle}
\usepackage{cmap} 
\usepackage[cp1251]{inputenc}
\usepackage[english,russian]{babel}
\usepackage{braket,amssymb,amsmath,array,commath,mathtext,wrapfig,tikz}
\usepackage[arrows=pgf-filled,version=3]{mhchem}
\usepackage{bm}
\usepackage{amsthm,amsfonts,amscd}
\usepackage{graphicx}
\usepackage{cite}
\usepackage{lscape}
\usepackage{geometry}
\usepackage{indentfirst}
\usepackage{afterpage,soul,ulem,dcolumn,changebar,longtable,hhline,multirow,makecell}
\usepackage[small,bf]{caption}
\makeatletter
\renewcommand{\@biblabel}[1]{#1.\hfil} 
\makeatother

\begin{document}
\renewcommand{\refname}{References}
\begin{center}
\textit{The eternal naked singularity formation in the case of gravitational collapse of generalized Vaydia spacetime} \\
\end{center}
\begin{center}
\textbf{Vitalii Vertogradov}
\end{center}
	\begin{center}
		Physics department, Herzen state Pedagogical University of Russia,
		
		48 Moika Emb., Saint Petersburg 191186, Russia

vdvertogradov@gmail.com\end{center}

\textbf{Summary:} 
In this paper, we consider the gravitational collapse of generalized Vaidya space-time when the matter satisfies the equation of the state either $P=0$ or $P=-\alpha \rho$, where $0 < \alpha < 1$. We show that in the case when type I of matter field is dust, then the apparent horizon will never appear, but there is now a family of null radial future-directed geodesics which terminated at the central singularity in the past. Also, we show that in the case of negative pressure, the result of the gravitational collapse might be the naked singularity and the apparent horizon appears and in very short time disappears again. In the case of the negative pressure, we show that the result of the gravitational collapse might be the eternal naked singularity.

\textbf{Keywords}: Gravitational collapse; Vaidya space-time; black hole; naked singularity. \\
\textbf{PACS numbers}: 04.70.—s, 04.70.Bw, 97.60.Lf

	\section{Introduction}

In 1939, Oppenheimer and Snyder~\cite{bib:open}  considered gravitational collapse of presureless homogeneous matter cloud. They showed that the result of such collapse is a black hole. The result of the gravitational collapse was believed to be only black hole for a long time. But then it was understood that during the gravitational collapse the singularity might be formed earlier than the apparent horizon(see for example~\cite{bib:joshi1, bib:joshi2}). The apparent horizon is the marginally trapped surface and it is the boundary of the trapped region. So if it is absent, then if there is a family of non-spacelike future-directed geodesics which terminate at the central singularity in the past then such singularity is visible and is called the naked singularity. If we consider the gravitational collapse of the inhomogeneous dust, then the result of such collapse might be the naked singularity. It is worth noting that this naked singularity is temporary and in a short period of time, it will be covered with the apparent horizon and we have a black hole as the result of the gravitational collapse.

In~\cite{bib:4} , Joshi and Goswami showed that the result of the gravitational collapse of a scalar field with small negative pressure might be the eternal naked singularity. We decided to consider the gravitational collapse of generalized Vaidya space-time~\cite{bib:5} in order to find out the possibility of the eternal naked singularity formation in this case. In~\cite{bib:6} Maharaj and Goswami showed that the result of the gravitational collapse of generalized Vaidya space-time depends on the mass function. Later in~\cite{bib:7}, the conditions for the mass function were given and according to these conditions, the result of the gravitational collapse (when the type I of the matter field satisfies the equation of the state $P=\alpha\rho$, where $ 0 < \alpha < \frac{1}{3}$) is the naked singularity. Here, we decided to consider two models of the gravitational collapse when type I of the matter field8 has either zero or negative pressure.

In Sec. 2, we consider the model of the gravitational collapse when type I of the matter field satisfies the equation of the state $P=0$. In Sec. 3, we consider the model with the negative pressure $P=-\alpha \rho$. In Sec. 4, we consider the model of the gravitational collapse when its result is the eternal naked singularity. Section 5 is the conclusion.

System of units $c=8\pi G=1$ will be used. Also, we denote $M'=\frac{dM(v,r)}{dr}$ and $\dot{M}=\frac{dM(v,r)}{dv}$.

\section{The Case When type I of the Matter Field is Dust}

The generalized Vaidya space-time is given by~\cite{bib:5}:
\begin{equation}
\begin{split}
ds^2=-(1-\frac{2M(v,r)}{r})dv^2+2dvdr+r^2d\Omega^2 \,, \\
d\Omega^2=d\theta+\sin^2 \theta d\varphi \,.
\end{split}
\end{equation}

where $M(v,r)$ is the mass function and $v$ is advanced Eddington time. The Einstein tensor has the following components:
\begin{equation}
\begin{split}
G_{00}=\frac{(2M-r)M''+2\dot{M}}{r^2}+\frac{1}{2}(1-\frac{2M}{r})(\frac{2M''}{r}+\frac{4M'}{r^2}) \,, \\
G_{01}=-\frac{2M'}{r^2}\,, G_{22}=-2rM'' \,, G_{33}=\sin^2 \theta G_{22} \,.
\end{split}
\end{equation}

We consider the mixture of I and II matter fields, where type I of the matter field is usual matter and type II is the null radiation. The energy-momentum tensor in this case is given by:
\begin{equation}
\begin{split}
T_{ik}=T^{(n)}_{ik}+T^{(m)}_{ik} \,, \\
T_{ik}^{(n)}=\mu l_il_k \,, \\
T^{(m)}_{ik}=\rho (l_in_k-n_il_k) \,, \\
\mu =\frac{2\dot{M}}{r^2} \,, \\
\rho =\frac{2M'}{r^2} \,, \\
P=-\frac{M''}{r} \,, \\
l_i=\Delta^0_i \,, n_i=\frac{1}{2}(1-\frac{2M}{r})\Delta^0_i-\Delta^1_i \,, \\
l^il_i=n^in_i=0 \,, l^in_i=-1 \,.
\end{split}
\end{equation}
Where $P$ is the pressure, $\rho$ is energy density and $L_i, N_i$ are 2 null vectors. 

Our model must be physically relevant. The energy-momentum tensor must satisfy energy conditions~\cite{bib:8, bib:9}. Weak, strong and dominant energy conditions demand:
\begin{equation}
\mu \geq 0 \,, \rho \geq P \geq 0 \,.
\end{equation}

From the fact that $P=0$ follows that $G_{22}=G_{33}=0$ and hence:
\begin{equation}
\begin{split}
M''=0 \,, \\
M(v,r)=c(v)+d(v)r \,.
\end{split}
\end{equation}

Then energy conditions require:
\begin{equation}
\dot{c}(v)+\dot{d}(v)r \geq 0 \,, d(v) \geq 0 \,.
\end{equation}

The equation of the apparent horizon is given by:
\begin{equation}
\begin{split}
1-\frac{2M}{r}=0 \,, \\
1-2d(v)-\frac{2c(v)}{r}=0 \,, \\
r=\frac{2c(v)}{1-2d(v)} \,.
\end{split}
\end{equation}

We see that  if $d(v)>\frac{1}{2}$ and if $\dot{d}(v) >0$ then the singularity will never be covered with the apparent horizon. 

If we proved that there is a family of non-spacelike future-directed geodesics which terminated at the central singularity in the past, then the result of such gravitational collapse would be the eternal naked singularity. We can consider null radial geodesics:
\begin{equation}
\frac{dv}{dr}=\frac{2r}{r-2c(v)-2d(v) r}\,.
\end{equation}

And a family of non-spacelike future-directed geodesics which terminate at the central singularity in the past exists if :
\begin{equation}
\lim\limits_{r \to 0, v\to 0} \frac{dv}{dr} > 0 \,.
\end{equation}

But the apparent horizon will never appear if $1-2d(v)<0$ and

\begin{equation}
\frac{dv}{dr}=\frac{r}{r(1-2d(v))_{<0}-2c(v)}< 0 \,.
\end{equation}

So in this model we have singularity which will never be covered with apparent horizon, but there is not a family radial null future-directed geodesics which terminated at the central singularity in the past. 

To explain the absence of the apparent horizon in this case we should consider the expansion $\theta$. In the case of generalized Vaydia spacetime  the expansion is given by:

\begin{equation}
\theta =e^{-\gamma} \frac{2}{r} \left ( 1-\frac{2c(v)+d(v)r}{r} \right ) \,.
\end{equation}
Where factor $e^{-\gamma}$ doesn't have any impact on sign $\theta$.

We know if $\theta$ is negative on a hypersurface then this hypersurface is a trapped one. 
In our case we have:
\begin{equation}
\begin{split}
\theta < 0 \to 1-\frac{2c(v)+d(v)r}{r} < 0 \,, \\
r(1-2d(v)) _{<0} - 2c(v) < 0 \,.
\end{split}
\end{equation}

The expression $1-2d(v)$ is less than zero because this is the condition of  the apparent horizon absence. 

And we see that in spite of the absence of the apparent horizon  the result of such collapse is not   the eternal naked singularity because  each hypersurface in this spacetime is trapped one. 

In this case all geodesics will terminate in the central singularity because $\theta < 0$ everywhere. It also means that there are not closed orbits in this spacetime. Hence this singularity is like the singularity of the future in Friedman model~\cite{bib:12}.

\section{The case of negative presure}

In this case we consider the equation of the state $P=-\frac{1}{2} \rho$.
Then the mass function is given by:

\begin{equation}
M(v,r)=c(v)+d(v)r^2 \,.
\end{equation}

In this model  the strong energy condition  is violated because $P<0$.  Weak and dominant energy conditions demand:
\begin{equation}
\dot{c}(v)+\dot{d}(v)r^2\geq 0\,, d(v) \geq 0 \,.
\end{equation}

From the fact that $M(0,0)=0$ we have $c(0)=0$.

Substituting this solution into the equation of the apparent horizon we find

\begin{equation}
\begin{split}
1-\frac{2M(v,r)}{r}=0 \,, \\
1-2d(v)r-\frac{2c(v)}{r}=0 \,.
\end{split}
\end{equation}

We see that when $c(0)=0$ and $d(0)=0$ then the apparent horizon is absent.
Solving this equation with respect to r we find:
\begin{equation} \label{eq:abs}
\begin{split}
2d(v)r^2-r+2c(v)=0 \,, \\
D=1-16c(v)d(v)\,.
\end{split}
\end{equation}

And we see the apparent horizon is absent when $c(v)d(v) > \frac{1}{16}$ . 
So we have the naked singularity formation when $v=0$ when the apparent horizon appears and then when 
$c(v)d(v) > \frac{1}{16}$ then the apparent horizon disappears again.

Now we must consider the question about the existence of non-spacelike future-directed geodesics which terminate at the singularity in the past.
For example, we can consider radial null geodesics
\begin{equation}
\frac{dv}{dr}=\frac{2r}{r-2c(v) - 2d(v)r^2} \,.
\end{equation}

 Let's denote $\lim\limits_{r \to 0 , v \to 0} \frac{dv}{dr}=x_0$, then we can put $d(v)=\xi v$ and $c(v)=\lambda v$  ( where $\lambda \,, \xi$ is real positive constants) and rewrite geodesics equation as:
\begin{equation}
\begin{split}
x_0=\frac{2}{1-2\lambda x_0} \,, \\
2\lambda x_0^2-x_0+2=0 \,, \\
x_{0 \pm} =\frac{1 \pm \sqrt{1-16\lambda}}{4} \,.
\end{split}
\end{equation}

The family non-spacelike future-directed geodesics exist if there is positive root of the equation above.  And $x_{0+}$ is positive, but we should constrain $\lambda$:
\begin{equation}
1-16\lambda \geq 0 \to \lambda \leq \frac{1}{16} \,.
\end{equation}
 So the result of such collapse might  be naked singularity.

If we look at the expression \eqref{eq:abs} then we will find that at the late stage of $v$ this expression has 2 positive roots. So we have three regions:
\begin{itemize}
\item Region I when $0\leq r \leq r_1$,
\item region II when $r_1 \leq r \leq r_2$,
\item region III when $r_2 \leq r < +\infty$.
\end{itemize}
 
Let's consider the expansion $\theta$:

\begin{equation} \label{eq:ex}
\theta =e^{-\gamma} \frac{2}{r} \left (1-\frac{2c(v)+2d(v)r^2}{r} \right ) \,. 
\end{equation}
Where as in previous section $e^{-\gamma}$ doesn't have any impact on sign of $\theta$. 

As we can see from \eqref{eq:ex}:
\begin{equation}
\begin{split}
\theta < 0 \to 0 \leq r < r _1 \,, \\
\theta > 0 \to r_1 < r < r_2 \,, \\
\theta < 0 \to r_2 <r < +\infty \,.
\end{split}
\end{equation}

This result is similar to Schwarzschild-de Sitter-Kottler Spacetime~\cite{bib:13} but with the equation of the state $P=-\frac{1}{2}\rho$. We can see that region I is a black hole solution with singularity in its center. If the observer is located in region II then he cannot get information from regions I and III.

In this case the result of the gravitational collapse might be the naked singularity, but it is temporary and in a very short period of time it will be covered with the apparent horizon. But this apparent horizon is white hole apparent horizon. Because if we look at \eqref{eq:ex} then we see that $\theta >0$ at $0<r<r_2$. In this case $r_1<0$. So we have in the region of spacetime where geodesics can't get into region $0<r<r_2$ from region $r_2<r<+\infty$. This white hole is shrinking. At the late stage of $v$ the expression \eqref{eq:ex} has 2 positive roots. And then two apparent horizons merge and we have as the result of such collapse - the singularity which is like the singularity of the future in Friedman model.

However, under other conditions the result of the gravitational collapse might be the eternal naked singularity. We will discuss it in the section below.

\section{The eternal naked singularity formation}

The naked singularity is eternal if there is a family of non-spacelike future-directed geodesics which terminate at the central singularity in the past, the apparent horizon  never appears and $\theta$ is positive  everywhere. 

Let's consider the case of negative  pressure when $\theta$ is always positive. The equation  of the state in this case is $P=-\alpha \rho$ where $1 \leq \alpha < 0$. $\alpha$ can't be more than 1 due to energy conditions  which demand $\rho \geq P \geq 0$.  The mass function in this case is given by:

\begin{equation}
M(v,r)=c(v)+d(v)r^{1+2\alpha} \,.
\end{equation}

The expansion in this case is given by:
\begin{equation} \label{eq:teta}
\theta =e^{-\gamma} \frac{2}{r} \left ( 1- \frac{2c(v)+2d(v)r^{1+2\alpha}}{r} \right ) \,.
\end{equation}
Here as in previous sections $e^{-\gamma}$ doesn't have any impact on sign of $\theta$.

Now we demand $\theta$ to be positive:
\begin{equation}\label{eq:neg}
\theta > 0 \to r > 2c(v)+2d(v)r^{1+2\alpha} \,.
\end{equation}

From \eqref{eq:neg} we see that we can satisfy this condition only if $c(v)\equiv 0$. Now we must constrain $\alpha$: $\alpha < 1$. Because if $\alpha =1$ then there is no a singularity. Finally we have:
\begin{equation} \label{eq:aah}
\begin{split}
\theta > 0 \to 1-2d(v)r^{2\alpha} > 0 \,, \\
r^{2\alpha}d(v) < \frac{1}{2} \,.
\end{split}
\end{equation}
From \eqref{eq:aah} we also see that there is no the apparent horizon because the condition to the apparent horizon formation is $\theta =0$ but we demand $\theta > 0$. Also from \eqref{eq:aah} we see that $d(v)$ must be very small and according to this the pressure must be very small because $P=-4\alpha (1+2\alpha )\frac{d(v)}{r^{2-2\alpha}}$. 

Now we should prove that there is a family of non-spacelike future-directed geodesics which terminate at the central singularity in the past. We consider null radial  geodesic which is given by:
\begin{equation}
\frac{dv}{dr}=\frac{2}{1-2d(v)r^{2\alpha}} \,.
\end{equation}
This family exists if:
\begin{equation}
\lim\limits_{v \to 0, r\to 0} \frac{dv}{dr} \geq 0 \,.
\end{equation}

But $\frac{dv}{dr}$ is always positive because $1-2d(v)r^{2\alpha}$ must be positive due to the condition $\theta > 0$.
So we have satisfied all three conditions and according to them the result of such collapse is the eternal  naked singularity.

Let's look at the  equation \eqref{eq:aah}. We can see that it has solution:
\begin{equation}
r_{aah}=\left (\frac{1}{2d(v)} \right )^{\frac{1}{1+2\alpha}} \,.
\end{equation}

We must match this solution to the exterior Schwarzschild        one.
So let $r_m$
 is the boundary of two spacetimes. Then we have two cases

\begin{enumerate}
\item $r_{aah} <r_m$,
\item $R_{aah} > r_m$.
\end{enumerate}

In the first case the result of gravitational collapse is a white hole because $\theta$ is positive at $0 \leq r < r_{aah}$. And there is a family of non-spacelike future-directed geodesics which terminate at the central singularity in the past.  
But we can also see that $\frac{dv}{dr}$ is always positive inside the white hole apparent horizon and there is not a geodesic which can get into white hole from the region outside the apparent horizon. But in this case it takes endless time to get to exterior observer from singularity.

In case II we match solutions at $r_m$ which is less than $r_{ah}$ and it takes   finite time to get to exterior observer from the central singularity. Now let's look at the expression of pressure, which is given by:

\begin{equation} \label{eq:res}
P=-2\alpha \left (1+2\alpha \right ) \frac{d(v)}{r^{1+2\alpha}} \,.
\end{equation}

From equations \eqref{eq:aah} and \eqref{eq:res} we can see that the less the pressure is the more $r_{ah}$ is. It means that the eternal naked singularity can be formed if the pressure is small enough.

Now let's consider the question about the strength of the singularity. According to ~\cite{bib:6} the singularity is gravitationally strong if:

\begin{equation} \label{eq:last}
\lim\limits_{\tau \to 0} \tau^2\psi =\frac{1}{4}x^2_0 \left (1 -\frac{4\alpha \mu^2}{2\alpha-1}x_0^{2\alpha-1}\right ) > 0\,,
\end{equation}

We know that:
\begin{equation}
\begin{split}
\mu =2\frac{\dot{c}+\dot{d}r^{1-2\alpha}}{r^2} \,, \\
\-1 <  \alpha < 0 \,, \\
 x_0=\lim\limits_{r\to 0, v\to 0} \frac{dv}{dr} \,.
\end{split}
\end{equation}

Now we suppose that:
\begin{equation}
\begin{split}
\alpha =-\frac{1}{2} \,, \\
d(v)=\xi v\,, \\
0 < \xi < 1 \,.
\end{split}
\end{equation}

In this case $d(v)$ satisfies weak and dominant energy conditions.

After simple calculations we obtain:

\begin{equation}
\lim\limits_{\tau \to 0}=1-\xi^2 \> 0 \,.
\end{equation}

Thus we have showed that the eternal singularity might be gravitationally strong.

\section{Conclusion}

In this paper, we have considered three models of the gravitational collapse of generalized Vaidya space-time. We found out that in the case when type I of the matter field satisfies the equation of the state $P=0$, then the apparent horizon might never appear but the singularity is not naked because there are trapped hypersurfaces. This singularity is similar to the singularity of the future in Friedman model.

When we have the equation of the state $P=-\alpha \rho$ then the result of gravitational collapse might be the naked singularity and the apparent horizon appears and in a very short period of time, it disappears again. But in this case, however, the naked singularity is not eternal.

Also, we consider the model of gravitational collapse when its result is the eternal naked singularity. But in this case, the pressure must be very small. Also, it is worth emphasizing that last two models violate strong energy condition like in the case of regular black holes.~\cite{bib:10, bib:11}

\textbf{Acknowledgment}
The author would like to thank Professor Pankaj Joshi for scientific discussion and this work was supported by RFBR Grant 18-02-00461.

\end{document}